\documentstyle[preprint,prd,aps,epsf,epsfig]{revtex}

\begin{document}
\draft
%
\title{  Limits of the Quantum Monte Carlo method
}
\author
{Z. N\'eda  and Z. Dezs\H{o}}
\address{Babe\c{s}-Bolyai University, Dept. of Theoretical Physics\\
str. Kog\u{a}lniceanu 1, RO-3400, Cluj-Napoca, Romania}

\maketitle
\centerline{\small (Last revised 5 Dec. 1999)}

\begin{abstract}

We consider the one-dimensional quantum-statistical problem of interacting
spin-less particles in an infinite deep potential valley and on a ring.
Several limits for the applicability of the quantum Monte Carlo methods
were revealed and discussed. We show the inapplicability of the
quantum Monte Carlo method for ring-like geometries, realize
an unphysical frustration for interacting fermions and a minus-sign
problem for interacting bosons.



\end{abstract}

\vspace{1cm}


\section{Introduction}

Quantum Monte Carlo (QMC) \cite{rev} methods are known to be powerful tools in
studying various quantum-statistical problems like quantum-spin models
\cite{q-spin},
Hubbard or t-J models \cite{hubbard},
quantum chromodynamical systems \cite{quark} or interacting
fermions in real space \cite{fermions}. It is believed that the method has great potentials
to handle analytically difficult and complex problems. The main idea of the
method is to transform a $d$ dimensional quantum-statistical problem in
$d+1$ dimensional classical one, by a procedure closely related to the
Path integral formalism of quantum mechanics \cite{path}. Due to the increased
dimensionality the method is computer-time consuming. Nowadays,
the accessible
and powerful modern computers makes the method more and more popular.

As an exercise for the applicability of QMC methods we studied the
quantum-statistical  problem of several short-range interacting quantum
particles in one-dimensional space.  During this exercise we found
however several technical and conceptual difficulties
which limits the applicability of
the method. The present paper intends to discuss the problem in this sense.

\section{The method}

In this section we review briefly the method of studying short-range
interacting quantum particles in real 1D spaces by the QMC method.
We will neglect the spin variables and consider idealized spin-less fermions
or bosons.

The time-independent Schr\"odinger equation for a single particle
with mass $m$  in 1D is:
\begin{equation}
\hat{H}\psi = -\frac{h^2}{8 \pi ^2 m} \Delta \psi +V(x) = E \psi.
\label{sch}
\end{equation}
Discretazing the problem on a lattice with sites  of
length $a$, the differential equation (\ref{sch}) can be written in the form
of
$L$ ($aL= l$ with $l$ the length of the considered space)
coupled linear equations
\begin{equation}
(-\frac{h^2}{8 \pi^2 m a^2})(\psi_{i+1}+\psi_{i-1}-2 \cdot \psi_i) + V_i \cdot
\psi_i = E\psi_i,
\label{diff}
\end{equation}
where $\psi_i$ and $V_i$ denotes the medium values in  box $i$ of the
lattice for the functions
$\psi(x)$ and
$V(x)$ respectively.
Introducing the notations
\begin{eqnarray}
t=\frac{h^2}{8 \pi^2 m a^2} \\
W_i=V_i+\frac{h^2}{4 \pi^2 m a^2},
\end{eqnarray}
equation (\ref{diff}) becomes:
\begin{equation}
-t\cdot (\psi_{i+1}+\psi_{i-1}) + W_i\psi_i=E\psi_i
\label{simp}
\end{equation}
We can write now this system of equations in a second quantized form,
using as state vectors the $n_i$ occupation number of the cells:
$\mid n_1, n_2,...,n_i,...,n_L>$.
We consider the  $\hat{c}_i^+$ creation,  $\hat{c}_i$ annihilation and
$\hat{n}_i=\hat{c}_i^+ \hat{c}_i$
occupation number operators defined for bosons as
\begin{eqnarray}
\hat{c}_i^+\mid n_1,n_2,...,n_i,...,n_L>=\sqrt{n_i+1}\mid n_1,n_2,...,n_i+1,...n_L>\\
\hat{c}_i\mid n_1, n_2,...,n_i,...,n_L>=\sqrt{n_i}\mid n_1,n_2,....n_i-1,...n_L>
\end{eqnarray}
and for fermions ($n_i=\{0, 1\}$) as
\begin{eqnarray}
\hat{c}_i^+\mid n_1, n_2,...,n_i,...,n_L>=(-1)^{s_i}(1-n_i) \mid n_1, n_2,...,n_i+1,...n_L>\\
\hat{c}_i\mid n_1, n_2,...,n_i,...,n_L>=(-1)^{s_i}n_i \mid n_1, n_2,...,n_i-1,...,n_L>,
\end{eqnarray}
where $s_i=\sum_{j=1}^{i-1} n_j$.
The wave function of the system becomes
\begin{equation}
\psi=\sum_{i=1}^L \psi_i \hat{c}_i^+\mid 0,0,,....,0>,
\end{equation}
and equations (\ref{simp}) can be written as:
\begin{equation}
\sum_{i=1}^L[-t(\hat{c}_{i+1}^+ \psi_i + \hat{c}_i^+ \psi_{i+1}) + W_i \hat{c}_i^+ \psi_i]
\mid 0,0,...,0>= \sum_{i=1}^L E \psi_i \hat{c}_i^+ \mid 0,0,...,0 >.
\end{equation}
The second quantized  form of the
Hamiltonian in the discretized space  will
be:
\begin{equation}
\hat{H}=-t\sum_i [\hat{c}_i^+\hat{c}_{i+1} + \hat{c}_{i+1}^+ \hat{c}_i] + \sum_i W_i \hat{n}_i .
\end{equation}
Considering more then one non-interactive particles in the same lattice
the Hamiltonian is unchanged. Interactions between
the particles will be introduced via extra terms like
\begin{equation}
\hat{H}_o=V_o \cdot \sum_i \hat{n}_i (\hat{n}_i-1),
\label{i1}
\end{equation}
or:
\begin{equation}
\hat{H}_1=V_1\cdot \sum_i \hat{n}_i \hat{n}_{i+1}.
\label{i2}
\end{equation}
Term (\ref{i1}) represent an on-site repulsion, and (\ref{i2})
the interaction between particles in neighboring cells.
Depending on the sign of $V_1$ this interaction can be attractive
or repulsive.

The partition function is
given by
\begin{equation}
Z= Tr(\exp{(-\beta \cdot \hat{H})})=\sum_{\{ n_i \} } <n_1,n_2,...,n_L\mid \exp{
(-\beta \cdot \hat{H})} \mid n_1,n_2,...,n_L>,
\end{equation}
where the sum is over all the possible combination of the
$n_i$ occupation numbers subject to the $\sum_{i=1}^L n_i = N$ ($N$ the
total number of particles) restriction and
$\beta=1/k_B T$.

For a practically successful realization of the MC simulation we have
to rewrite the $Z$ partition function  in a form in which
calculation of the $P$ transition probabilities are easy when only
a few $n_i$ occupation numbers have been changed.
We will sketch how is possible this.

Denoting by
$V_i$ the interaction terms of the forms $V_o \hat{n}_i(\hat{n}_i-1)$ and,
$V_1 \hat{n}_i \hat{n}_{i+1}$,
we separate our Hamiltonian in two commuting parts
$\hat{H}_a$ and $\hat{H}_b$.
\begin{eqnarray}
\hat{H}_a=(-t\hat{c}_1^+\hat{c}_2-t\hat{c}_2^+\hat{c}_1+\frac{V_1}{2} + \frac{V_2}{2})+ \nonumber \\
+(-t\hat{c}_3^+c_4-t\hat{c}_4^+\hat{c}_3+\frac{V_3}{2} + \frac{V_4}{2}) + ..... \nonumber \\
...+(-t\hat{c}_{L-1}^+\hat{c}_{L}- t\hat{c}_{L}^+\hat{c}_{L-1} + \frac{V_{L-1}}{2} + \frac{V_L}{2})=
\nonumber \\
=\hat{H}_1+\hat{H}_3+... + \hat{H}_{L-1},
\end{eqnarray}
with:
\begin{equation}
\hat{H}_i=-t\hat{c}_i^+\hat{c}_{i+1}-t\hat{c}_{i+1}^+\hat{c}_i+ \frac{V_i}{2} + \frac{V_{i+1}}{2}.
\end{equation}
In a similar way $\hat{H}_b=\hat{H}_2+\hat{H}_4+.....+\hat{H}_L$.
One will observe immediately that the terms inside
$\hat{H}_a$ and $\hat{H}_b$ commute, and so:
\begin{eqnarray}
< \{ n_i \} \mid \exp{(-\beta \hat{H}_a)} \mid \{ n_j \} > = < \{ n_i \} \mid \exp {
(- \beta \hat{H}_1)} \mid \{ n_j \} > \cdot  \nonumber \\
\cdot < \{ n_i \} \mid \exp{(-\beta \hat{H}_3)} \mid
\{ n_j \} >........
\cdot < \{ n_i \} \mid \hat{H}_{L-1} \mid \{ n_j \}>.
\end{eqnarray}
The same equation is valid for $\hat{H}_b$.

Ideal it would be to write the partition function as
product of exponentials, each of them containing  one
$\hat{H}_i$ term. In this way the calculation of the transition probabilities
when only a few occupation numbers are changed would be easier.
To achieve this we need to write
$\exp{[-\beta (\hat{H}_a+\hat{H}_b)]}$ as: $\exp{(-\beta \hat{H}_a)}
\cdot \exp{(-\beta \hat{H}_b)}$.
Unfortunately this equation is not valid anymore, because $\hat{H}_a$ and $\hat{H}_b$ do not
commute.

However, for two $\hat{A}$ and $\hat{B}$ sufficiently small
operators one can use the approximation:
\begin{equation}
e^{\hat{A}}\cdot e^{\hat{B}}= e^{\hat{A}+\hat{B}+0.5*[\hat{A},\hat{B}]}
\approx e^{\hat{A}+\hat{B}}.
\label{t-s}
\end{equation}
Considering now $M$ a large integer, so that
$\frac{\beta
\hat{H}_a}{M}$ and $\frac{\beta \hat{H}_b}{M}$ is small enough, we are able to
use the (\ref{t-s})  approximation in the form proposed by
Trotter and Suzuki  \cite{t-s}:
\begin{eqnarray}
Z=\sum_{\{ n_i \} } < \{ n_i \} \mid (\exp{[-\frac{\beta}{M}\cdot
(\hat{H}_a+\hat{H}_b)]})^M \mid \{ n_i
\} > \approx \nonumber \\
\approx \sum_{ \{ n_i \} } < \{ n_i \} \mid
(\exp{[-\frac{\beta}{M}\cdot \hat{H}_a]} \cdot
\exp{[-\frac{\beta}{M}\cdot \hat{H}_b]})^M \mid \{ n_i \} >
\label{TS}
\end{eqnarray}

In order to write  $Z$ as a product,
which in a MC simulation can be considered as transition
probabilities, we insert between each exponentials in (\ref{TS}) a complete
set of states:
\begin{equation}
\sum_{ \{ n_{i,j} \} } \mid n_{1,j},n_{2,j},....,n_{L,j}> <
n_{1,j},n_{2,j},....,n_{L,j} \mid.
\end{equation}

Because each of these
$2M -1$ sets represent an independent sum over all possible
states, they will be indexed with a new label,
$j$, different from the spatial one, $i$.
The initial trace contribute also to the independent set of states,
so the $j$ index can take
$2M$ values.
Remembering that $\hat{H}_i$ acts only on the states
$n_i$ and $n_{i+1}$ we can write:
\begin{equation}
Z\approx \sum_{ \{n_{i,j } \}} \prod_{j=1}^{2M} \prod_{i=1}^L P_{i,
j }.
\end{equation}
When $j$ and $i$ are both even or odd the $P_{i, j}$
factors are calculable as
\begin{equation}
P_{i,j}=<n_{i,j},n_{i+1,j}\mid e^{(-\Delta \tau \cdot \hat{H}_i
)} \mid n_{i,j+1}, n_{i+1,j+1}>,
\label{P}
\end{equation}
and $P_{i,j}=1$ for all other choices
(we used the notation $\Delta \tau=\frac{\beta}{M}$).

Because both $i$ and the $j$ indices label occupation numbers,
by using the Trotter-Suzuki (T-S) approximation we created a two-dimensional
lattice from our real one-dimensional one. Every box of the original
lattice is multiplied $2M$ times, and all the $n_{i,j}$
sets ($j=\overline{1,2M}$) are independent.
Due to the original trace the $j$ index must satisfy the periodical boundary
conditions:
$n_{i,2M+1}=n_{i,1}$.

We are able now to consider our quantum-statistical problem as a classical one,
characterized by the
\begin{equation}
Z\approx \sum_{ \{ n_{i,j} \} } \exp{(-\beta \sum_{i,j} e_{i,j})}
\label{Z}
\end{equation}
partition function. In equation (\ref{Z}) we used the
$-\beta e_{i,j}=ln P_{i,j}$ notation and every $e_{i,j}$
is calculable from the four $n_{i,j}$, $n_{i+1,j
}$, $n_{i,j+1}$ and $n_{i+1,j+1}$ occupation numbers.

We have transformed our quantum-statistical
problem in a classical one, considered in a space with increased
dimensionality. The dimension indexed by $i$ is called spatial, and
the one indexed by $j$ the Trotter or imaginary time direction.
The presented method is totally equivalent
with the Path Integral [6] formulation of the quantum-mechanical problem.
(Each two lines in the Trotter direction corresponding to a time
interval
$\Delta \tau$  in the Path Integral formalism.)

The Monte Carlo simulation will follow now the known Metropolis algorithm.
In the transformed, (\ref{Z}) problem we have four-site
interactions between the neighboring sites
$n_{i,j}$, $n_{i+1,j}$, $n_{i,j+1}$  and
$n_{i+1,j+1}$, with the values of $i$ and $j$
both even or odd (otherwise
$e_{i,j}=0$).
In the  $i$ and $j$ space
this interaction can be represented by a check-board pattern
(Fig. 1), with interactions around the
dark plaquettes.

In many cases the
$P_{i,j}$ probabilities (\ref{P}) are zero,
and so the corresponding state is realized with zero probability.
The condition for
$P_{i,j}\ne 0$, is:
\begin{equation}
n_{i,j}+n_{i+1,j}=n_{i,j+1}+n_{i+1,j+1}.
\label{cond}
\end{equation}

One will realize immediately that in MC simulations
the changes leading to acceptable configurations are quite limited.
To save precious computer time and calculations, we must know
from the beginning which changes will give
nonzero transition probabilities. The acceptable changes
(leading to nonzero
transition probabilities ) will be:
\begin{itemize}
\item the occupation numbers from  the left side of a white
plaquette are increased by unity, and the values from the right
side decreased by unity
\item the occupation numbers from the left side of a white
plaquette are decreased by unity, and the values from the right site
increased by unity
\end{itemize}
(These changes can be done only if the occupation
numbers will not become negative, or for fermions if they are not
bigger than one.)

It is obvious that by satisfying initially the
(\ref{cond}) conditions for the whole system, and by
considering only the above mentioned changes in the occupation
numbers, the algorithm will lead to configurations keeping (\ref{cond}).

The Metropolis MC algorithm will be now as follows:
\begin{itemize}
\item we consider an initial configuration so that condition (\ref{cond})
holds for the whole lattice
\item we change the values of the occupation numbers around a white
plaquette in the way described earlier
\item we calculate the change in the total energy of the system
as
$\Delta E= \sum \Delta e_{i,j}$, the sum being done
on the neighboring dark plaquettes
\item we accept this change with a probability
\begin{equation}
P=\frac{\exp{(-\beta \Delta E)}}{\exp{(-\beta \Delta E)} + 1}
\end{equation}
\item we continue the algorithm until thermodynamic equilibrium is approached
\item we collect
periodically the relevant data.
\end{itemize}

We define one MC step as
$2ML$ trials of changing the configuration of
the system.

\section{Determining the relevant physical quantities}

The relevant physical quantities are determined after the equilibrium
dynamic is approached, by averaging over many MC steps (usually of order
$10^5$). The $E$ average energy and the $C$ heat-capacity is
computed by using the known equations:
\begin{eqnarray}
& E=-\frac{\partial \ln Z}{\partial \beta} \\
& C=\frac{1}{kT^2} \frac{\partial^2 \ln Z}{\partial \beta^2}
\end{eqnarray}
Keeping in mind that the $e_{ij}$ factors are also functions of $\beta$
from (\ref{Z}) we get:
\begin{eqnarray}
& E=\left< \sum\limits _{i,j} e_{ij} \right>+ \left< \beta \sum\limits_{ij}
\frac{\partial
e_{ij}}{\partial \beta} \right> \label{E}\\
& \nonumber C=\frac{1}{kT^2}\left( \left< \left(\sum\limits_{i,j} e_{ij} +
\beta \sum\limits_{ij}
\frac{\partial e_{ij}}{\partial \beta} \right)^2 \right> -
\left( \left< \sum\limits_{i,j} e_{ij} + \beta \sum\limits_{i,j}
\frac{\partial e_{ij}}
{\partial \beta} \right> \right)^2 \right) - \\
& -\frac{1}{kT^2}  \left< 2 \sum\limits_{i,j} \frac{\partial e_{ij}}
{\partial \beta} + \beta
\sum\limits_{i,j} \frac{\partial^2 e_{ij}}{\partial \beta^2} \right>
\label{C}
\end{eqnarray}
In the above formula all the sums are over the dark plaquettes of the
lattice and the averaging  is an ensemble average,
which in our QMC scheme is an average over different
MC steps.

\section{Calculation of the $e_{i, j}$ factors}

  We will calculate the $e_{ij}$ "energies" both for one particle
and for many interacting fermions and bosons.
For the later case we will assume that the interaction potential
between the particles in the same cell
is two times stronger
than the one between two particles in the nearest cells.
We will also assume that there is no external potential energy,
thus $W_i=2t$.

\subsection{One quantum particle}

The Hamiltonian of the particle depends on the kinetic energy only:
\begin{equation}
\hat{H}=-t\sum_{i=1}^L(\hat{c}_i^+\hat{c}_{i+1}+\hat{c}_{i+1}^+\hat{c}_i)+2t\cdot\sum_{i=1}^L n_i
\end{equation}
We use the symmetric form:
\begin{eqnarray}
 \hat{H}_i &=& -t\cdot(\hat{c}_i^+\hat{c}_{i+1}+\hat{c}_{i+1}^+\hat{c}_i)+
t\cdot \hat{n}_i+t\cdot \hat{n}_{i+1}
\ \  \mbox{where} \ \ i=\overline{1,L} \\
 \hat{H}&=&\sum_{i}^L \hat{H}_i=-t \sum_{i=1}^L \hat{O}_i   \\
  \hat{O}_i &=& (\hat{c}_i^+\hat{c}_{i+1}+\hat{c}_{i+1}^+\hat{c}_i)-\hat{n}_i
-\hat{n}_{i+1}, \ \ \mbox{where} \ \ i=\overline{1,L}.
\end{eqnarray}

One can write the $e_{ij}$ values as:
\begin{equation}
e_{ij}
=-\frac{1}{\beta} \cdot \ln \left( \sum_{k=0}^{\infty}
                     \left( \frac{\beta t}{M} \right)^k \cdot \frac{1}{k!}
                     <n_{i,j},n_{i+1,j}|\hat{O}_i^k |n_{i,j+1},n_{i+1,j+1}>\right)  \label{eij} \\
\end{equation}
$e_{i,j}$ depends only on the
 occupation numbers $n_{i,j}$,
$n_{i+1,j}$, $n_{i,j+1}$, $n_{i+1,j+1}$, thus  \\
$e_{i,j}=e_{i,j}(n_{i,j},n_{i+1,j},n_{i,j+1},n_{i+1,j+1})$.

For one particle
\begin{equation}
\sum_{i=1}^L n_{i,j}=1,
\end{equation}
and  only the terms with  $n_{i,j}+n_{i+1,j}=n_{i,j+1}+n_{i+1,j+1}$
can be different from zero: $e_{ij}(0,0,0,0)$, $e_{ij}(1,0,1,0)$,
$e_{ij}(0,1,0,1)$, $e_{ij}(1,0,0,1)$ and $e_{ij}(0,1,1,0)$.
The $e_{ij}(0,0,0,0)$ term is zero, and by using the fact that the
$\hat{O}_i$ operators are hermitian it is easy to prove:
\begin{eqnarray}
e_{i,j}(1,0,1,0) &=& e_{i,j}(0,1,0,1) \nonumber \\
e_{i,j}(1,0,0,1) &=& e_{i,j}(0,1,1,0).
\end{eqnarray}
Using the immediate
\begin{eqnarray}
\hat{O}_i|1,0> &=& 1 \cdot |0,1>-1 \cdot |1,0> \nonumber \\
\hat{O}_i|0,1> &=& 1 \cdot |1,0>-1 \cdot |0,1>,
\end{eqnarray}
equations, the two nontrivial $e_{ij}$ terms can be calculated by
a recursion formula:
\begin{eqnarray}
<1,0|\hat{O}_i^k|1,0> &=&
<1,0|\hat{O}_i^{k-1}|0,1>-<1,0|\hat{O}_i^{k-1}|1,0> \nonumber  \\
<1,0|\hat{O}_i^k|0,1> &=&
<1,0|\hat{O}_i^{k-1}|1,0>-<1,0|\hat{O}_i^{k-1}|0,1>.
\end{eqnarray}
Let us denote by  $a_k=<1,0|\hat{O}_i^k|1,0>$ and $b_k=<1,0|\hat{O}_i^k|0,1>$,
and rewrite the above recursion:
\begin{eqnarray}
a_k &=& b_{k-1}-a_{k-1} \nonumber \\
b_k &=& a_{k-1}-b_{k-1}.
\end{eqnarray}
Because $<1,0|1,0>=1$ and
$<1,0|0,1>=0$ the first terms of these series are $a_0=1$ and $b_0=0$.
For $k>0$ we get the analytical forms:
\begin{eqnarray} \label{rekur}
a_k &=& (-1)^k \cdot 2^{k-1} \nonumber \\
b_k &=& (-1)^{k+1} \cdot 2^{k-1}.
\end{eqnarray}
Substituting these into the
expression of the $e_{ij}$ terms (\ref{eij})  we get:
\begin{equation}
e_{i,j}(1,0,1,0)=
\frac{1}{\beta}\left(\frac{\beta t}{M}-\ln\left(
\cosh\left(\frac{\beta t}{M}\right)\right)\right),
\end{equation}
\begin{equation}
e_{i,j}(1,0,0,1)=
\frac{1}{\beta}\left(\frac{\beta t}{M}-\ln\left(
\sinh\left(\frac{\beta t}{M}\right)\right)\right).
\end{equation}

\subsection{Many interacting fermions}

The Hamiltonian in this case is:
\begin{equation}
\hat{H}=-t\sum_{j=1}^L\left( \hat{c}_{j+1}^+\hat{c}_j+\hat{c}_j^+
\hat{c}_{j+1}\right)+2t\sum_{j=1}^L \hat{c}_j^+\hat{c}_j+V_1\sum_{j=1}^L
\hat{n}_j\hat{n}_{j+1}.
\end{equation}

Let us assume that the particles have  charge q and there are rejective
forces between them.
The interaction is only between the nearest cells, because spin-less fermions
can not be simultaneously in the same cell. (This is the manner we impose
the antisymmetric wave-function for fermions).
The interaction potential is $V_1$
\begin{equation}
V_1=\frac{1}{4\pi\varepsilon_0}\cdot\frac{q^2}{a},
\end{equation}
where $a$ is the width of the cells.

The Hamiltonian terms are
\begin{equation}
\hat{H}_i=-t(\hat{c}_i^+\hat{c}_{i+1} + \hat{c}_{i+1}^+\hat{c}_{i}
-\hat{n}_i-\hat{n}_{i+1}-2b\hat{n}_i\cdot \hat{n}_{i+1}),  \nonumber
\end{equation}
with  $b=\frac{V_1}{t}$.
The occupation numbers for spin-less fermions can be
either  zero or one, thus  $e_{ij}$ has three
different values: $e_{ij}(1,0,1,0)$, $e_{ij}(1,0,0,1)$
and $e_{ij}(1,1,1,1)$.
The first two  were already calculated.

For the calculation of $e_{ij}(1,1,1,1)$ we show that
\begin{equation} \label{rek11}
\hat{O}_i^k|1,1>=-2(1+b)\hat{O}_i^{k-1}|1,1>,
\end{equation}
and get:
\begin{equation}
<1,1|\hat{O}_i^k|1,1>=[-2(1+b)]^k.
\end{equation}
Substituting this in the expression of the $e_{ij}$ energy:
\begin{equation}
e_{ij}(1,1,1,1)=
\frac{2t(1+b)}{M}.
\end{equation}

\subsection{Many interacting bosons}

In this case more particles can be in the same cell, so
the interaction between particles in the same cell has to be
considered too.
The Hamiltonian has the following form:
\begin{equation}
\hat{H}=-t\sum_{j=1}^L \left(\hat{c}_{j+1}^+\hat{c}_j+\hat{c}_j^+
\hat{c}_{j+1}\right)+2t\sum_{j=1}^L \hat{c}_j^+\hat{c}_j+V_1\sum_{j=1}^L
\hat{n}_j\hat{n}_{j+1}+V_0\cdot \sum_{j=1}^L \hat{n}_j(\hat{n}_j-1)\nonumber.
\end{equation}
Let us assume $V_0=2V_1$. In our standard notations:
\begin{equation}
\hat{H}_i=-t[\hat{c}_i^+\hat{c}_{i+1}+\hat{c}_{i+1}^+\hat{c}_i-
\hat{n}_i-\hat{n}_{i+1}-2b\hat{n}_i\hat{n}_{i+1}-b\hat{n}_i(\hat{n}_i-1)-
b\hat{n}_{i+1}(\hat{n}_{i+1}-1)].
\end{equation}

In case of N bosons we have two conditions:
\begin{eqnarray}  \label{fel}
\sum_{i=1}^Ln_{i,j} &=& N, \nonumber \\
n_{i,j}+n_{i+1,j} &=& n_{i,j+1}+n_{i+1,j+1}.
\end{eqnarray}

Due to the $i\leftrightarrow i+1$ and $j\leftrightarrow j+1$ invariance of
the $e_{ij}$ values
we consider only the dark squares with the biggest occupation
number in the left bottom corner.
For simplicity we introduce the following notation
\begin{equation}
<n_{i,j},n_{i+1,j}|\hat{O}_i^k|n_{i,j+1},n_{i+1,j+1}>=
(n_{i,j},n_{i+1,j},n_{i,j+1},n_{i+1,j+1})_k,
\end{equation}
and calculate the recurrence formula for a given $n=n_{i,j}+n_{i+1,j}=n_{i,j+1}+n_{i+1,j+1}$
occupation numbers ($n\in[0,N]$).
If $n$ is odd and $n>1$:
\[\left(\frac{n+2l+1}{2},\frac{n-2l-1}{2},\frac{n+2k+1}{2},\frac{n-2k-1}{2}
\right)_{k}= \]
       \[         \sqrt{\frac{(n+2k+3)(n-2k-1)}{4}}
          \left(\frac{n+2l+3}{2},\frac{n-2l-3}{2},
          \frac{n+2k+1}{2},\frac{n-2k-1}{2}\right)_{k-1}+ \]
       \[          \sqrt{\frac{(n+2k+1)(n-2k+1)}{4}}
           \left(\frac{n+2l+1}{2},\frac{n-2l-1}{2},
          \frac{n+2k+1}{2},\frac{n-2k-1}{2}\right)_{k-1}-\]

\begin{equation}\label{impar}
            n(1+b(n-1))
        \left(\frac{n+2l+1}{2},
\frac{n-2l-1}{2},\frac{n+2k+1}{2},\frac{n-2k-1}{2}
          \right)_k.
\end{equation}
The possible values of $l$ and $k$  are
$l\in\left[0,\frac{n-1}{2}\right]$ and
$k\in[-l-1,l]$.

\newpage

If n is even
\[\left(\frac{n}{2}+l,\frac{n}{2}-l,\frac{n}{2}+k,\frac{n}{2}-k
\right)_{k}= \]
       \[         \sqrt{\frac{n+2k+2}{2} \frac{n-2k}{2}}
          \left(\frac{n+2l}{2},\frac{n-2l}{2},
          \frac{n+2k+2}{2},\frac{n-2k-2}{2}\right)_{k-1}+ \]
       \[          \sqrt{\frac{n+2k}{2}\frac{n-2k+2}{2}}
           \left(\frac{n+2l}{2},\frac{n-2l}{2},
          \frac{n+2k-2}{2},\frac{n-2k+2}{2}\right)_{k-1}-\]
\begin{equation} \label{par}
            n (1+b(n-1))
         \left(\frac{n+2l}{2},\frac{n-2l}{2},\frac{n+2k}{2},\frac{n-2k}{2}
          \right)_k,
\end{equation}
where $l\in\left[0,\frac{n}{2}\right]$ and $k\in[-l,l]$.

This equations carries all the necessary information for
calculating  numerically the  $e_{ij}$ energies.

\section{Test for one particle in an infinite deep potential valley}

We verified our QMC algorithm by considering first one quantum particle
(an electron) in an infinite deep potential valley
and in contact with a heat-bath at
temperature $T$. This simple quantum-statistical problem is easily
computable. The energy levels are given by the well-known formula:
\begin{equation}
E_n=\frac{h^2 n^2}{8 m L^2 a^2}, \: \: (n=1,2, ....),
\end{equation}
($a L$  is the width of the valley, and was taken $3 \cdot 10^{-9}m$)
The expectation value of the energy is numerically computable as:
\begin{equation}
<E>_t=\frac{\sum_{n=1}^{\infty} E_n exp(-\beta E_n)}
{\sum_{n=1}^{\infty} exp(-\beta E_n)}.
\end{equation}
One can also easily compute theoretically the heat-capacity, of
the system by numerically derivating $<E>_t$ as a function of $T$.

Considering a series of simulations with $L=12$ and $p=\beta t/M=0.1$
we got the $<E>_{QMC}$ values plotted with dots in Fig. 2a. This result is
presented in comparison with the $<E>_t(T)$ theoretical curve.
On Fig. 2b we plotted  together the theoretically computed heat-capacity
(dashed line), the one obtained by QMC applying
formula
(\ref{C}) (filled circles) and the one obtained from the
\begin{equation}
C=\frac{\Delta <E>_{QMC} }{\Delta T},
\label{CDT}
\end{equation}
formula (empty circles).
The correspondence is satisfactory.

\section{Problem with one quantum particle on a ring}

After the promising results for the infinite deep potential valley it is quite
frustrating to realize that in this case we cannot solve the original
quantum-statistical problem. In the QMC method imposing periodic
boundary conditions in the spatial direction will define a problem for
a quasi-free particle in constant potential field. The reason for this is
that in the QMC formalism we loose the wave-function and remain only with
the occupation probabilities. In this manner  it is impossible to
impose the necessary closing conditions for the wave-function and it's
derivative. From the symmetry of the problem we get only the condition that
the occupation numbers have translational invariance. This defines
a problem for a free quantum particle (i.e infinite wide potential valley).
We expect in this manner to find
\begin{eqnarray}
& <E>_t=\frac{kT}{2}, \\
& C=\frac{k}{2},
\end{eqnarray}
which are the known results for a free particle in contact with a heat-bath
at temperature $T$. In contrast, by imposing the ring-like geometry in
the quantum-mechanical problem, the closing condition for the wave-function
would yield the
\begin{equation}
E_n=\frac{h^2n^2}{2ma^2L^2}, \:  \: (n=1,2,3......),
\label{En}
\end{equation}
energy levels and a different $<E>_t'(T)$ dependence.

The previous predictions are totally supported by our simulation results.
We considered a quantum particle with the mass of an electron, $aL=3\cdot
10^{-9}m$ and $L=18$.
On Fig.~3a we plotted the $<E>_{QMC}$ simulation points together with the
expected $<E>=kT/2$ curve (continuous line) and the solution with
the discrete (\ref{En}) energy levels
for the quantum states (dashed line).
The simulation results are convincingly supporting the
free quantum particle solution. The heat-capacity obtained from
formula (\ref{C}) fluctuates around the expected $k/2$ value (Fig.~3b).

Through this simple exercise one can immediately realize a
very important limitation for the QMC methods.
By loosing the $\Psi$ wave-function, closing or boundary conditions
for $\Psi$ and it's derivative are replaced with continuity
conditions for $\mid \Psi \mid^2$ and particle flux $\vec{j}$. The
two problems however are not equivalent, and important differences
can be obtained for several problems.

\section{Minus-sign problem for interacting bosons}

For bosons where the occupation numbers inside one cell can be arbitrary high
an immediate difficulty arises when calculating the (\ref{P})
$P_{ij}$ transition
probabilities. For large occupation numbers the $P_{ij}$ values become
negative, and thus the QMC method is inapplicable. This problem is similar
with the well-known minus-sign problem for the Hubbard model \cite{minus}.
The origin
of this non-physical situation is the used Trotter-Suzuki approximation
and thus the finite value of $p=\beta t/M$. As the value of $p$ is lower more
bosons can be considered inside one spatial cell without getting minus
sign for $P_{ij}$. On Fig.~4   as a function of $p$ we plotted the maximal
boson number per cell for which the $P_{ij}$ factors are all positive.
(We considered here free bosons with the mass and charge of an electron
interacting
with a repulsive Coulomb potential, and $a=1.66\cdot 10^{-10}m$.)
From this figure we learn that in the limit $p\rightarrow 0$ the minus-sign
problem disappear. Unfortunately in this limit the $M$ value (lattice size
in the imaginary Trotter direction) has to be infinite. The practical solution
is of course by fixing the maximal number of bosons in the simulation. This
leads to a maximal value of $p$, and thus a finite $M$ value. In this manner
the minus-sign problem will be eliminated. The technical problem we will
face now is that even for small number of bosons the required $M$ value is
large and this limits seriously the applicability of the QMC method for the
proposed problem.

\section{Unphysical frustration for interacting fermions}

In MC-type simulations the usual method for calculating the heat-capacity
is by the use of the fluctuation-dissipation theorem. In our case
this leads to equation (\ref{C}), and we get a $C_1$ value.
An alternative method for calculating the
heat-capacity ($C_2$) could be by the definition of this quantity,
i.e. by using equation (\ref{CDT}).

Considering the problem of interacting fermions (spin-less electrons)
on a ring-like geometry
we proposed to determine the specific heat of the system both as a function
of temperature and particle number in the system. We considered
repulsive Coulomb interactions, $a=1.66 \cdot 10^{-10}m$, and
$L=18$ . In order to get confidence
in our calculations we used both of the methods. The obtained results were
in agreement for low and high fermion number, and totally contradictory for
medium occupation density. The logarithm of the ratio
($C_{1}/C_2$ )
of the two specific heat is plotted in Fig. 5. Verifying and testing
the result in several aspects we got confidence in the obtained
differences. The explanation of this strange behavior lies
in the space discretization and the way of handling the exclusion principle
in QMC simulations. The combined effect of this two introduces a
non-physical frustration in the system. This
frustration is maximal for half-filled cells  and leads to
results similar with the ones obtained in spin glasses below the freezing
temperature. In MC simulations for spin glasses the
difference between the specific
heat determined in the two different manner is known \cite{spin-glass}, and
indicates that thermal equilibrium is not completely reached.
We encounter a similar situation also here.

To better understand our previous
arguments let us follow what is happening in our 2D lattice following
the particles paths. Each fermion is represented by a
continuous series of $1$ values in the Trotter direction which defines
the "path" (or the "world-line") of the particle. For fermions these paths
cannot intersect each other and can change maximally one cell
size for 1 step in the Trotter dimension. A possible path configuration
for two fermions is shown as an example in Fig.~6. If we consider more than
two fermions, due to the introduced space discretization and the restrictive
condition for fermion numbers inside one cell many paths become frustrated.
A typical situation for three fermions is shown in Fig.~7. In this setup
the path for the fermion in the middle is frustrated in our QMC
algorithm. By a no energy change this path can fluctuate between
the two neighbors.
The energetically stable position is in the middle, but due to
our space discretization there is no cell there.
We expect that this effect is maximal around the half-filling
and disappears for less than three fermions or  for complete filling.
This prediction is in agreement with the observed results.

The solution for
eliminating this unphysical frustration is  by considering a fine space
discretization where the frustration is minimized.
Unfortunately this will increase again
the lattice size and create  technical problems.

\section{Conclusions}

Considering the quantum statistical problem of interacting
particles on a ring-like geometry and in an infinite deep potential
valley we realized several limits of the QMC method.

\begin{itemize}
\item We loose the concept of wave-function and the method
becomes  inappropriate for ring-like geometries.
Imposing periodic boundary conditions will lead to the problem
of particles in a spatially constant potential.
\item For interacting bosons
we found that due to the considered
Trotter-Suzuki transformation a minus-sign problem appears.
As the Trotter dimension is increased (the Trotter-Suzuki approximation
is improved) more and more bosons per
spatial cell can be considered without getting into the minus-sign problem.
This problem can be eliminated by increasing the lattice size, which leads
to increased computational time.
\item For interacting fermions the combined effect of
spatial discretization and the
application of the exclusion principle creates an unphysical
frustration in the system. In this way thermal equilibrium is hard to
reach and problems similar to the one obtained in MC methods for
spin-glasses are encountered. The effect of this unphysical frustration can be
minimized by further increasing the spatial cell numbers, which creates
again technical difficulties.
\end{itemize}

In conclusion, for a successful application of the QMC method on the
proposed problem one has to keep in mind and eliminate the above limitations.


\vspace{1cm}




\begin{figure}[htp]
\epsfig{figure=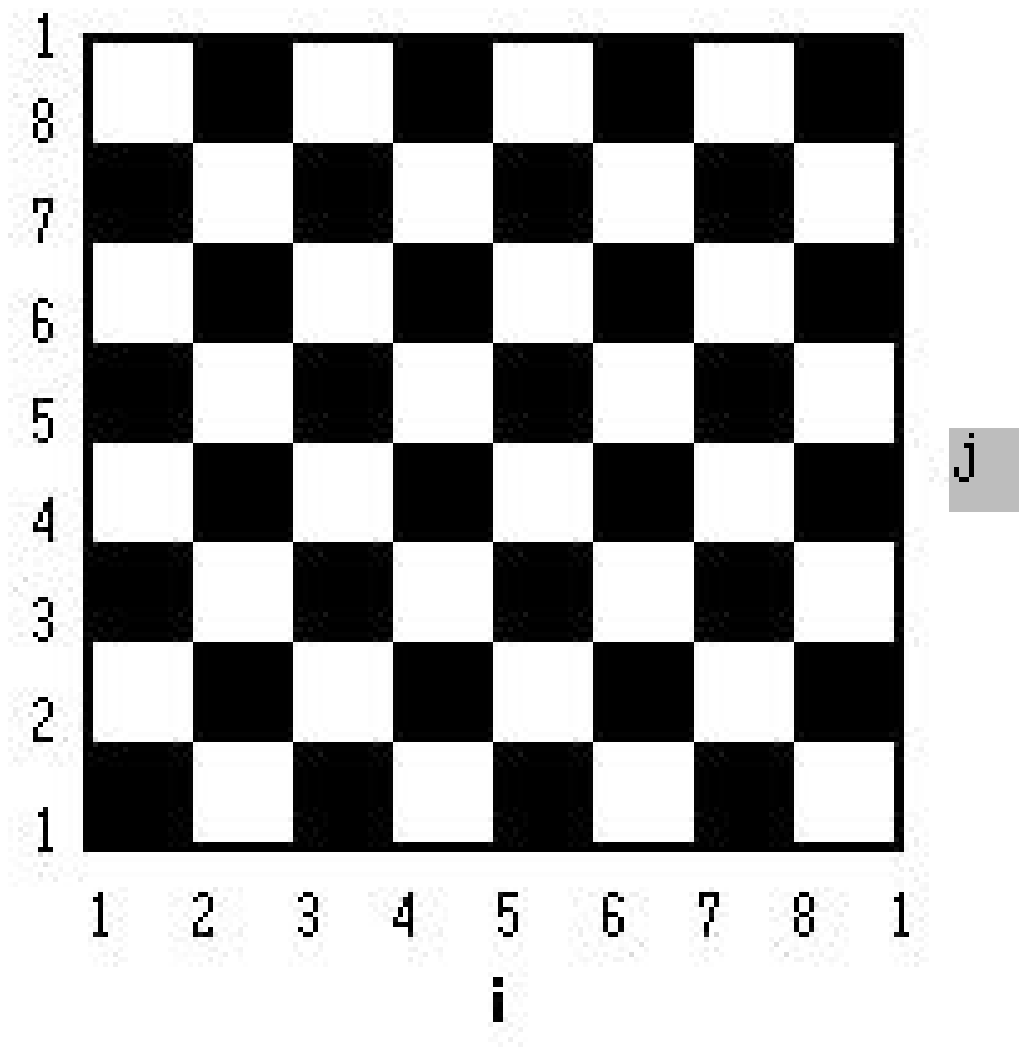,height=2.5in,width=2.5in,angle=-0}
\caption{ The characteristic chess-board lattice for the
QMC simulations}
\label{fig1}
\end{figure}


\begin{figure}[htp]
\epsfig{figure=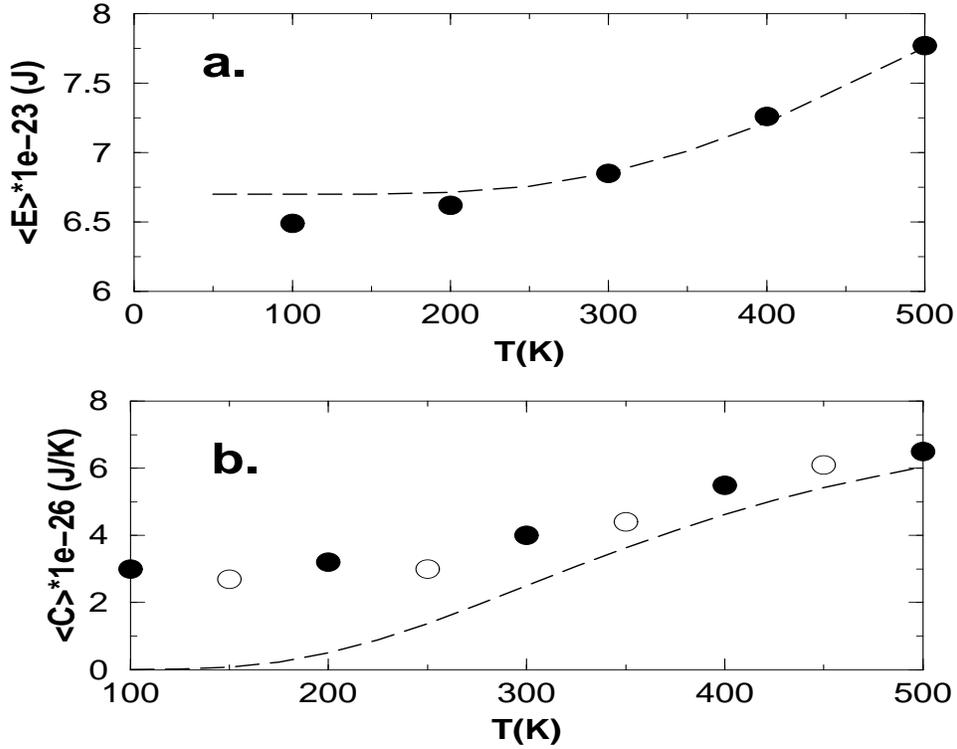,height=4.0in,width=5.0in,angle=-0}
\caption{ QMC results in comparison with the exact theoretical
results for an electron in a 1D potential valley with width
$a L=3\cdot 10^{-9}m$. Fig.2a
presents the energy values of the system as a function of the
heat-bath temperature. The dashed curve is theoretical, the
filled circles are QMC results. Fig.2b presents the heat-capacity
of the system as a function of the temperature. The dashed curve is
the exact theoretical result, filled circles are QMC results
obtained by (\ref{C}), and the empty circles are results obtained from
(\ref{CDT}). ($L=12$ and
$p=0.1$) }
\label{fig2}
\end{figure}


\begin{figure}[htp]
\epsfig{figure=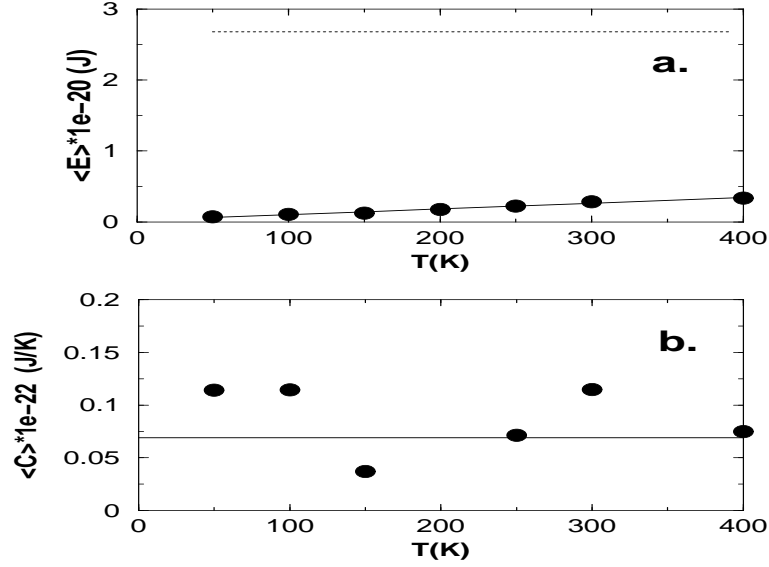,height=3.0in,width=4.0in,angle=-0}
\caption{QMC results in comparison with theoretical results
for an electron  on a ring (length of the ring: $3\cdot 10^{-9}m$).
Fig.~3a show the energy
values as a function of the heat-bath temperature. Filled circles
are QMC simulation results,  dashed curve is the desired result
for the right quantum-mechanical problem, and the continuous curve
presents the theoretical result for a  free
electron (i.e. infinite large ring). On Fig.~3b we present the
heat-capacity of the system as a function of the temperature.
Circles are QMC results, obtained by (\ref{C}), and the line
represent the theoretical $k/2$ value for a free particle.
($L=18$ and $p=0.2$)}
\label{fig3}
\end{figure}


\begin{figure}[htp]
\epsfig{figure=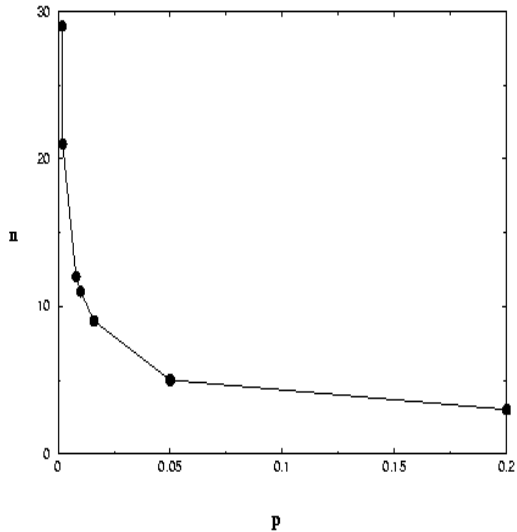,height=3.5in,width=3.5in,angle=0}
\caption{Maximal bosons number per cell ($n$), for which the minus-sign
problem does not appear as a function of the $p$ parameter. We
considered bosons with the mass and charge of the electron, interacting with
repulsive Coulomb potential. ($a=1.66 \cdot 10^{-10}m$ and $L=18$)}
\label{fig4}
\end{figure}


\begin{figure}[htp]
\epsfig{figure=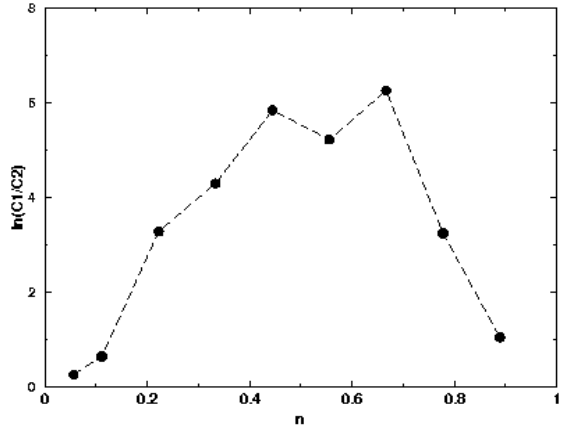,height=3.0in,width=4.0in,angle=-0}
\caption{ Logarithm of the ratio for the heat-capacity calculated
by two different ways ($C_1$ and $C_2$), as a function of
particle number per cell.
(Interacting fermions with mass and charge of an electron,  Coulomb
potential, $aL=3 \cdot 10^{-9}m$ and $L=18$.)
}
\label{fig5}
\end{figure}


\begin{figure}[htp]
\epsfig{figure=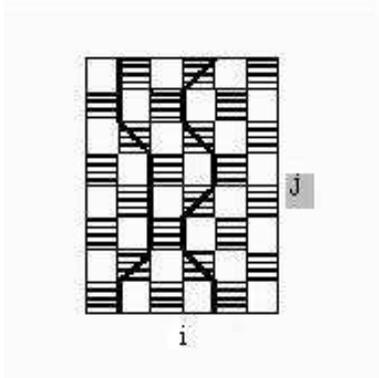,height=2in,width=2.0in,angle=-0}
\caption{A specific path configuration for two fermions}
\label{fig6}
\end{figure}


\begin{figure}[htp]
\epsfig{figure=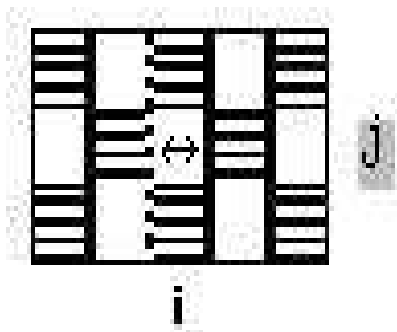,height=2.0in,width=3.0in,angle=-0}
\caption{ Example of frustration  for three fermions}
\label{fig7}
\end{figure}

\end{document}